\documentclass{aastex}

\def \inte {$INTEGRAL$}
\def \xmm {$XMM$-$Newton$}
\def \sax {$BeppoSAX$}
\def \sw {$Swift$}

\def \src {IGR\,J16328-4726}

\def \hcm {\hbox {\ifmmode $ atom cm$^{-2}\else atom cm$^{-2}$\fi}}

\def \arcsec {\hbox{$^{\prime\prime}$}}

\def \apj {ApJ}

\def \aap {A\&A}

\newcommand{\be}{\begin{equation}}

\newcommand{\ee}{\end{equation}}

\begin{document}

\title{The \xmm\ \thanks{XMM Newton is an ESA science mission with instruments and contributions directly funded by ESA member States and the USA (NASA)} and \inte\ \thanks{INTEGRAL is an ESA project with instruments
 and science data centre funded by ESA member states
(especially the PI countries: Denmark, France, Germany, Italy, Switzerland, Spain),
 Czech Republic and Poland, and with the participation of Russia and the USA.} observations  of the
supergiant fast X--ray transient \src}

\author{M. Fiocchi\altaffilmark{1}, A. Bazzano\altaffilmark{1}, L. Natalucci\altaffilmark{1}, P. Ubertini\altaffilmark{1}, V. Sguera\altaffilmark{2}, A. J. Bird\altaffilmark{3}, C. M. Boon \altaffilmark{3}, P. Persi\altaffilmark{1}, L. Piro\altaffilmark{1}    } 

\altaffiltext{1}{INAF, Istituto di Astrofisica e Planetologia Spaziali, Via Fosso del Cavaliere 100,   I-00133 Roma,  Italy }
\altaffiltext{2}{ INAF, Istituto di Astrofisica Spaziale e Fisica Cosmica, Via Gobetti 101, I-40129 Bologna, Italy}
\altaffiltext{3}{School of Physics and Astronomy, University of Southampton, University Road, Southampton, SO17 1BJ, UK}

\begin{abstract}
The accretion mechanism producing the short flares observed from the 
Supergiant Fast X-ray Transients (SFXT) is still highly debated and forms a major part in our attempts to place these X-ray binaries in the wider context of the High Mass X-ray Binaries. 

We report on a 216 ks \inte\/ observation of the SFXT \src\  (August 24-27, 2014) simultaneous with two fixed-time observations with \xmm\ (33ks and 20ks)
performed around the putative periastron passage, in order to investigate the accretion regime and the wind properties during this orbital phase.\\
During these observations, the source has shown luminosity variations, from $\rm \sim4\times10^{34} erg~s^{-1}$ to $\rm \sim10^{36} erg~s^{-1}$,
linked to spectral properties changes. 
The soft X-ray continuum is well modeled by a power law with a photon index varying from $\sim$1.2  up to $\sim$1.7  and with high values of the column density in the range $\rm \sim 2-4\times10^{23} cm^{-2}$.
 We report on the presence of iron lines at $\sim$ 6.8-7.1 keV suggesting that the X-ray flux is produced by accretion of matter from the companion wind characterized by density and temperature inhomogeneities. 

\end{abstract}

%\keywords{editorials, notices --- 
%miscellaneous --- catalogs --- surveys}

\section{Introduction} \label{sec:intro}
\src\/ was first reported as an unidentified transient source in the third \inte\/ IBIS/ISGRI  survey
with a flux of 4 mCrab and 3.2 mCrab in the energy range 20-40 keV and 40-100 keV, respectively (Bird et al. 2007, Bodaghee et al. 2007). 

The  \sw\/ XRT follow-up observations performed during a flare (on 2009 June 10) showed that the X-ray spectrum of the source
was, at that time, well described by an absorbed power law model with a $N_{\rm H}$$\simeq$8$\times$10$^{22}$~cm$^{-2}$ in excess of the expected Galactic value  and a photon index $\Gamma$$\simeq$0.56 (Grupe et al. 2009).

On the basis of its transient and recurrent nature, 
its short and intense flares and a dynamic range of $\sim10^2$,
this source has been classified as a candidate SFXT (Fiocchi et al. 2010).
An orbital period corresponding to $\sim$ 10 days has been derived by Corbet et al. (2010) making use of 
\sw\//BAT data. 
\src\/ was  observed  by \xmm\/ on  2011 February  20 (corresponding to an orbital phase of $\sim$0.1) for  a  total  exposure  time  of $\sim$22  ks (Bozzo et al. 2012). The analysis of these data showed a flux variation of a factor $\sim$10  without 
significant variation 
of the spectral parameters (N$_H$ and $\Gamma$).  The average spectrum was well fitted with an absorbed power law model with a column density of $\rm \sim 17.5\times10^{22}cm^{-2}$, a photon index of $\sim 1.5$
and unabsorbed 2-10 keV flux of $\rm 1.7\times10^{-11}erg~cm^{-2}s^{-1}$.
The source was also  within the field of view of \sax\/ in 1998: the MECS X-ray data showed a frequent microactivity typical of the intermediate state of SFXT and a weak flare with a duration of $\sim$4.6 ks (Fiocchi et al. 2013). 
 During these observations the photon index of the power law model remained constant while the absorption column density was highly variable, spanning from $\sim$3 to $\sim$20 $\rm \times10^{22}cm^{-2}$ across the transition from the low emission level
 ($\rm F_{2-10keV}\sim3\times10^{-12}erg~cm^{-2}~s^{-1}$) to the peak of the flare ($\rm F_{2-10keV}\sim10^{-10}erg~cm^{-2}~s^{-1}$). 
Romano et al. 2013 reported on the spectral analysis of a flare that occurred on 2009 June 10, and has observed with the \sw\//XRT instrument. During the brightest X-ray emission 
(unabsorbed flux of   $\rm 4.2\times10^{-10}erg~cm^{-2}~s^{-1}$) the photon index was $\sim$0.65 and the column density was $\rm \sim$9$\times$10$^{22}$~cm$^{-2}$ in excess of the Galactic one. 
IR observations allowed  confirmation of 
the nature of the companion as a O8I spectral type star  (Coleiro et al. 2013) and determined the source distance of 7.2$\pm$0.3 kpc (Persi et al. 2015). 
At soft X-ray energies a long term monitoring (2011-2013) with \sw\//XRT allowed a detailed study of the emission outside the bright outbursts, identifying two low emission levels, both well described with a power law model:
the first has a photon index of $\sim1.35$ and a column density of $\rm \sim13.6\times$10$^{22}$ cm$^{-2}$  at an observed flux of $\rm F_{2-10keV}\sim16\times10^{-12}erg~cm^{-2}~s^{-1}$, 
while the second one is fitted with a photon index
of $\sim0.3$ and a column density of $\rm \sim1.5\times$10$^{22}$ cm$^{-2}$ at an observed flux of $\rm F_{2-10keV}\sim1.1\times10^{-12} erg~cm^{-2}~s^{-1}$.
 These observations allowed an estimate of the lower limit of the dynamic range in this source of $\sim$750 (Romano et al. 2014b).\\
 This source has been reported as a SFXT characterized by an intermediate orbital period and a low flux variability in the review of Walter et al. (2015).

Outbursts from \src\/ usually occur near the periastron passage, at a restricted phase range of its orbital period (10.068$\pm$0.002\ days, Fiocchi et al. 2013), allowing the use of fixed-time observations.
In this paper we report on the spectral results for two \xmm\/ (Jansen et al. 2001) observations, performed quasi-simultaneously with a long \inte\/ (Winkler et al. 2003) 
observation performed at periastron (phase = 0.5).

 \section{Observations and Data Reduction\label{dataredu}}

%%%%%%%%%%%%%%%%%%%%%%%%%%%%%%%%%%%%%%%%%%%%%%%%%%%%%%%%%%%%%%%%%%%%%%%% 
\begin{figure}
\begin{center}
\centerline{\includegraphics[scale=0.65, angle=-90]{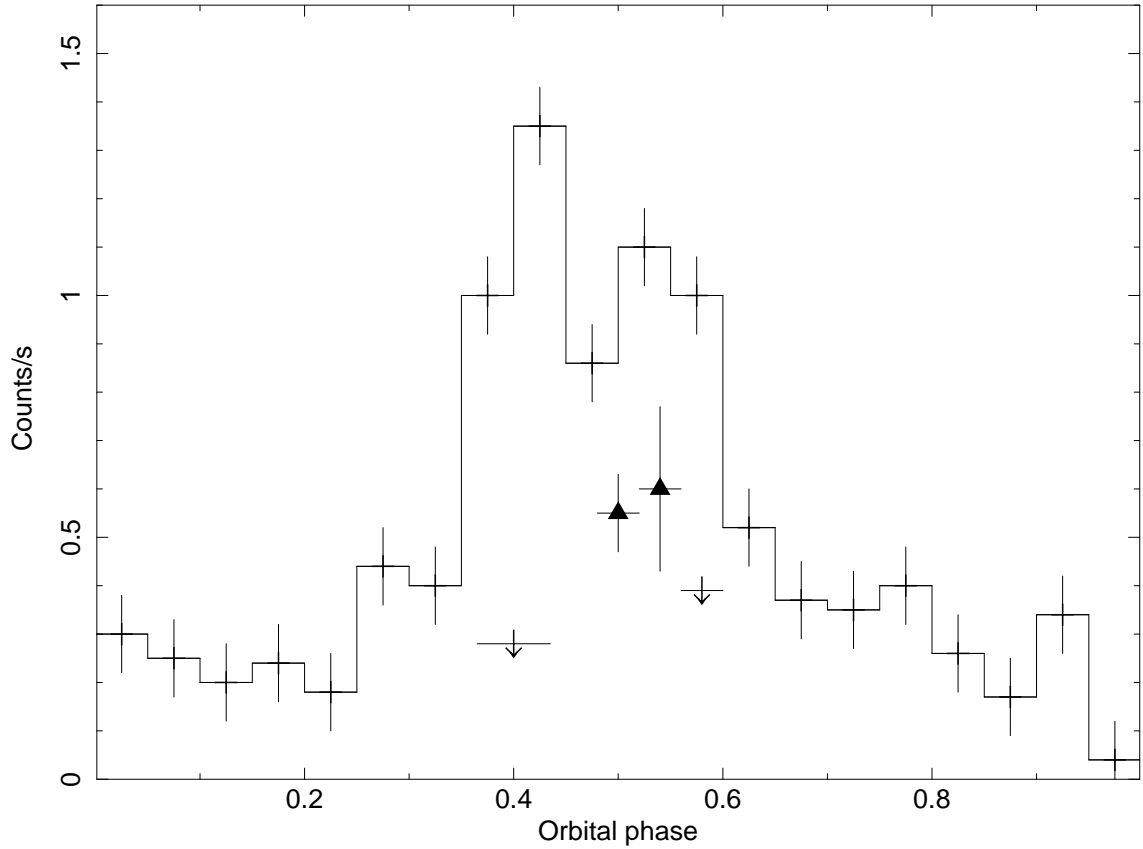}}
\caption{The (23-50 keV) IBIS/ISGRI intensity of \src\
during revolutions 1448 and 1449 (triangles and upper limits) superimposed on the phase-folded light curve (crosses), constructed using the best orbital period determination of 10.068 days and a zero-phase ephemeris of MJD 52651.164 
(from Fiocchi et al. 2013).
 } \label{fig:lc2}
\end{center}
\end{figure}
%%%%%%%%%%%%%%%%%%%%%%%%%%%%%%%%%%%%%%%%%%%%%%%%%%%%%%%%%%%%%%%%%%%%%%%%

%%%%%%%%%%%%%%%%%%%%%%%%%%%%%%%%%%%%%%%%%%%%%%%%%%%%%%%%%%%%%%%%%%%%%%%% 
\begin{figure*}
\begin{center}
\centerline{\includegraphics[width=19cm,angle=0]{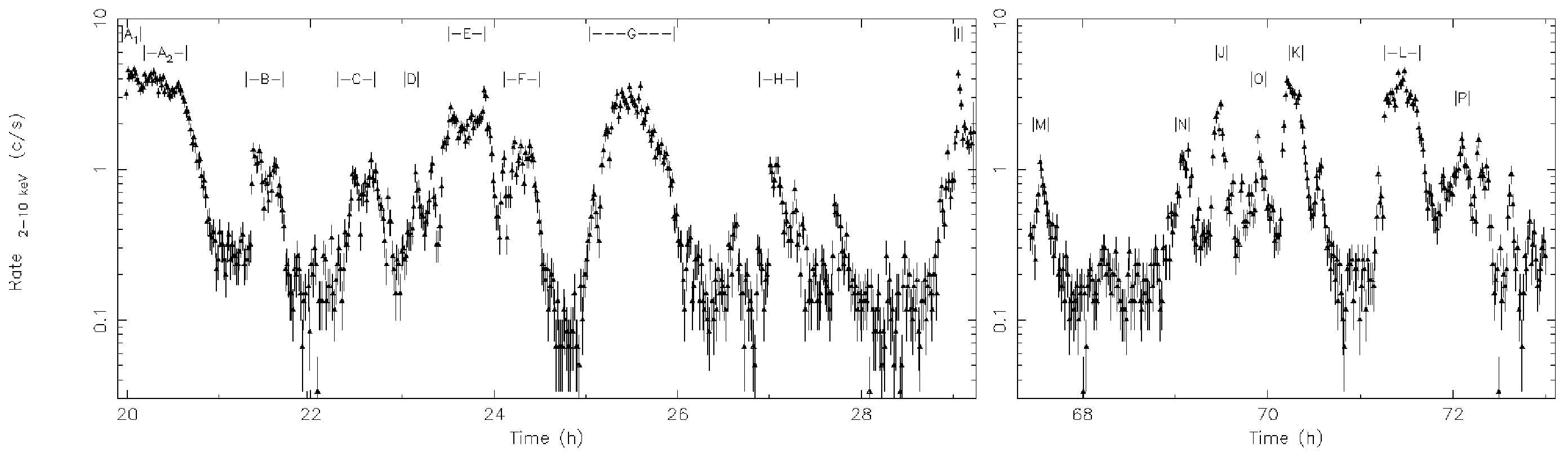}}
\caption{EPIC PN, background-subtracted light curves of \src\
in the 2--10 keV energy range. The bin size is 60~s and the
time axis is in UTC hours from the start time of 56893 19:59:29.57 MJD. Letters indicate the time intervals used for selected spectral states (see text for details). 
 The left panel shows the first \xmm\/ observation at orbital phase $\sim\/$0.4 and the right panel the second  \xmm\/ observation at orbital phase $\sim\/$0.6.
 } \label{fig:lc3}
\end{center}
\end{figure*}
%%%%%%%%%%%%%%%%%%%%%%%%%%%%%%%%%%%%%%%%%%%%%%%%%%%%%%%%%%%%%%%%%%%%%%%%
The \inte\/ observation commenced on 2014-08-24T11:38:42 (revolution 1448) and ended 2014-08-27T07:36:18 (revolution 1449).
The \xmm\ observations were performed on 2014-08-24T19:57:54 (UTC) for  $\sim$\/33ks (orbital phases $\sim$0.4)
and on  2014-08-26T19:24:52 (UTC) for $\sim$\/20ks (orbital phases $\sim$0.6). 

The INTEGRAL/IBIS (Ubertini et al. 2003) 
data are processed using the Off-line Scientific Analysis
(OSA v10.2) software released by the
\inte\/ Scientific Data Centre (Courvoisier et al. 2003). \\
Light curves and images of the source are extracted in the 23--50 keV energy band.\\
These runs were performed with the AVES cluster, designed to optimize performances and disk storage
need to the INTEGRAL data analysis by Federici et al. (2010).\\

\xmm\/ data were processed using version 14.0 of the Science Analysis
Software (SAS).
During both observations, EPIC MOS1 and PN operated in Full Frame Imaging mode and  EPIC MOS2 in Partial Frame Imaging mode, 
filter used was the medium thickness filter for both observations.
Calibrated events are filtered using patterns 0-4 for the PN and 0-12 for both MOS.
Extraction radii of 40\arcsec\ were  used for the
source events for both the PN and MOS cameras. 
Background counts were extracted from source free regions, 
in the same temporal intervals. 
Response and ancillary matrix files
were generated using the SAS tasks {\sc rmfgen} and {\sc arfgen}.
All spectra were binned with a minimum of 20 counts per bin.\\
All spectral
uncertainties and upper-limits are given at 90\% confidence for
one parameter of interest.\\

  	%%%%%%%%%%%%%%%%%%%%%%%%%%%%%%%%%%%%%%%%%%%%%%%%%%%%%
  	\section{Analysis and Results\label{result}}
  	%%%%%%%%%%%%%%%%%%%%%%%%%%%%%%%%%%%%%%%%%%%%%%%%%%%%%

\src\/ was not detected with IBIS in the entire observation (216 ks on-source) 
at a significance level greater than 4$\sigma$.  The 3$\sigma$ upper limit in the 23-50 keV energy range is $\rm \sim 3\times10^{-10} erg~cm^{-2} s^{-1}$, corresponding to a luminosity $L\rm \sim 2\times10^{36} erg~s^{-1}$, 
using a distance of 7.2 kpc (Persi et al. 2015).

Due to the transient nature of \src\/ 
we produced four mosaics (23-50 keV) with a shorter exposure time: for revolution 1448 we extracted one mosaic during the first $\sim$ 18 hours of observation (T$_{start}^{1}$=56893.49 MJD), while for 
revolution 1449 we created three mosaics covering three periods of $\sim$ 13.3 hours each, with T$_{start}^{2}$=56894.71 MJD, T$_{start}^{3}$=56895.23 MJD and T$_{start}^{4}$=56895.74 MJD.\\
Intensities for each mosaic are reported in Fig.~\ref{fig:lc2} (triangles and upper limits) superimposed on the phase-folded light curve (from Fig.~1 of Fiocchi et al. 2013). 
This plot shows an increased activity around phase 0.5, most probably associated to the transit near periastron,
although no strong flare was detected ($\rm \leq 2\times10^{36} erg~s^{-1}$). 
Consequently we noted that the source was significantly detected during phase 0.50-0.55 at 6.0 sigma with a flux corresponding to $\rm \sim 3\times10^{-11} erg~cm^{-2} s^{-1}$ ( $\rm L_X\sim 2\times10^{35} erg~s^{-1}$), 
while the first and last points 
of Fig.~\ref{fig:lc2} are 3$\sigma$ upper limits. We summed the data in the period in which the source was detected, obtaining a spectrum with a net exposure time of $\sim$60ks. With the \inte\/ 
short observation we are not able to constrain the physical spectral parameters and we report here the (23-50 keV) flux  of $\rm (2.8\pm1.2)\times10^{-11} erg~cm^{-2} s^{-1}$  with 
a photon index value fixed to 2.0 ($\chi^2/d.o.f.=6.5/5$).  
We note that IBIS data in the phase $\sim$0.45-0.50 are not available because of the non visibility period between one INTEGRAL revolution and the next. % (see Fig.~\ref{fig:lc1}). 
Unfortunately the \xmm\/ observations were performed 
during two periods in which \src\/ was not detected with IBIS, preventing a spectral study using simultaneous data. 

The EPIC PN background-subtracted light curves of \src\,
 in the 2-10 keV energy range are shown in Fig.~\ref{fig:lc3}, at $\sim$0.4 (left panel) and $\sim$0.6 (right panel) orbital phase.

%
%%%%%%%%%%%%%%%%%%%%%%%%%%%%%%%%%%%%%%%%%%%%%%%%%%%%%%%%%%%%%%%%%%%%%%%%
\begin{figure}
\begin{center}
\includegraphics[height=13cm,angle=-90]{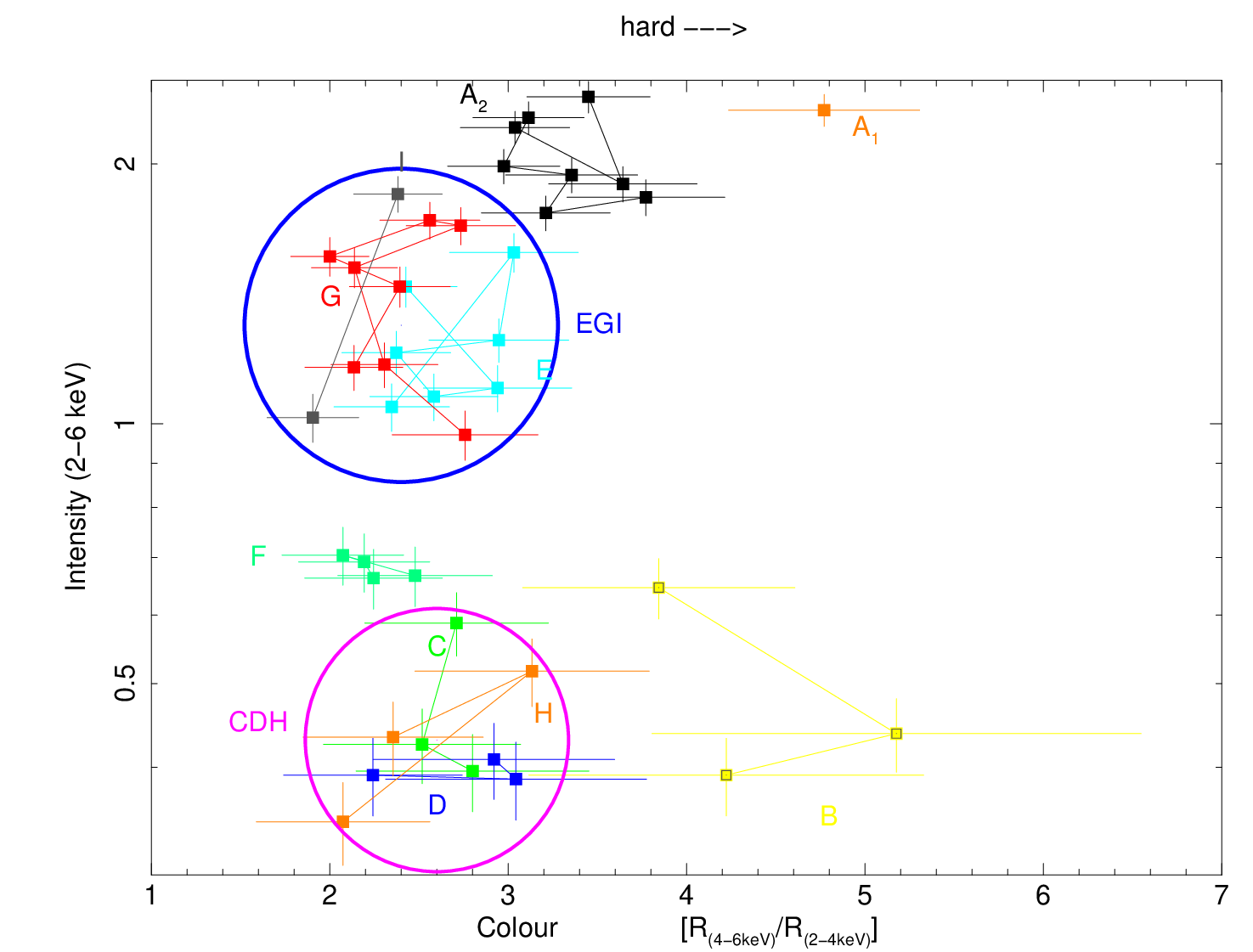} \\
\includegraphics[height=13cm,angle=-90]{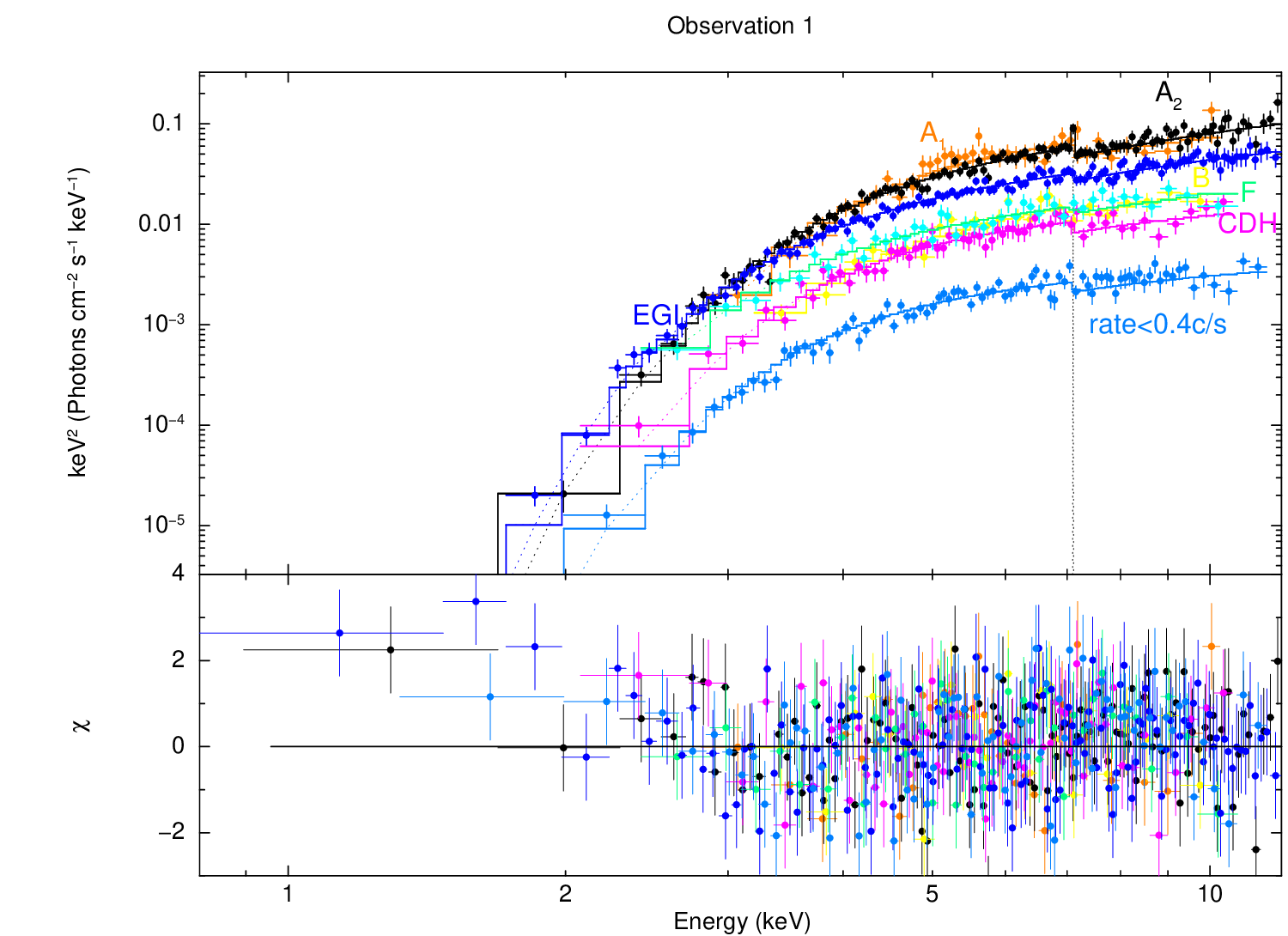} \\
\end{center}
\caption{Top panel: Hardness ratios $R=Rates_{4-6 keV}/Rates_{2-4 keV}$ for time intervals corresponding to flares displayed in Fig.~\ref{fig:lc3} plotted against the sum $S=Rates_{4-6 keV}/Rates_{2-4 keV}$ for the first observation.
Large circles indicate the combined data from flares with the similar flux level and hardness. Bottom panel: unfolded spectra and the model in $E^2f(E)$  and residuals of the following spectral states: A$_1$ in orange, A$_2$ in black, 
EGI in blue, F in green, B in yellow, CDH in magenta,  state with rate lower than 0.4 c/s in light blue, fitted with the best model described in the text. 
}
\label{fig:fluxratio1}
\end{figure}
%%%%%%%%%%%%%%%%%%%%%%%%%%%%%%%%%%%%%%%%%%%%%%%%%%%%%%%%%%%%%%%%%%%%%%%%

\begin{figure}
\begin{center}
\includegraphics[height=14cm,angle=-90]{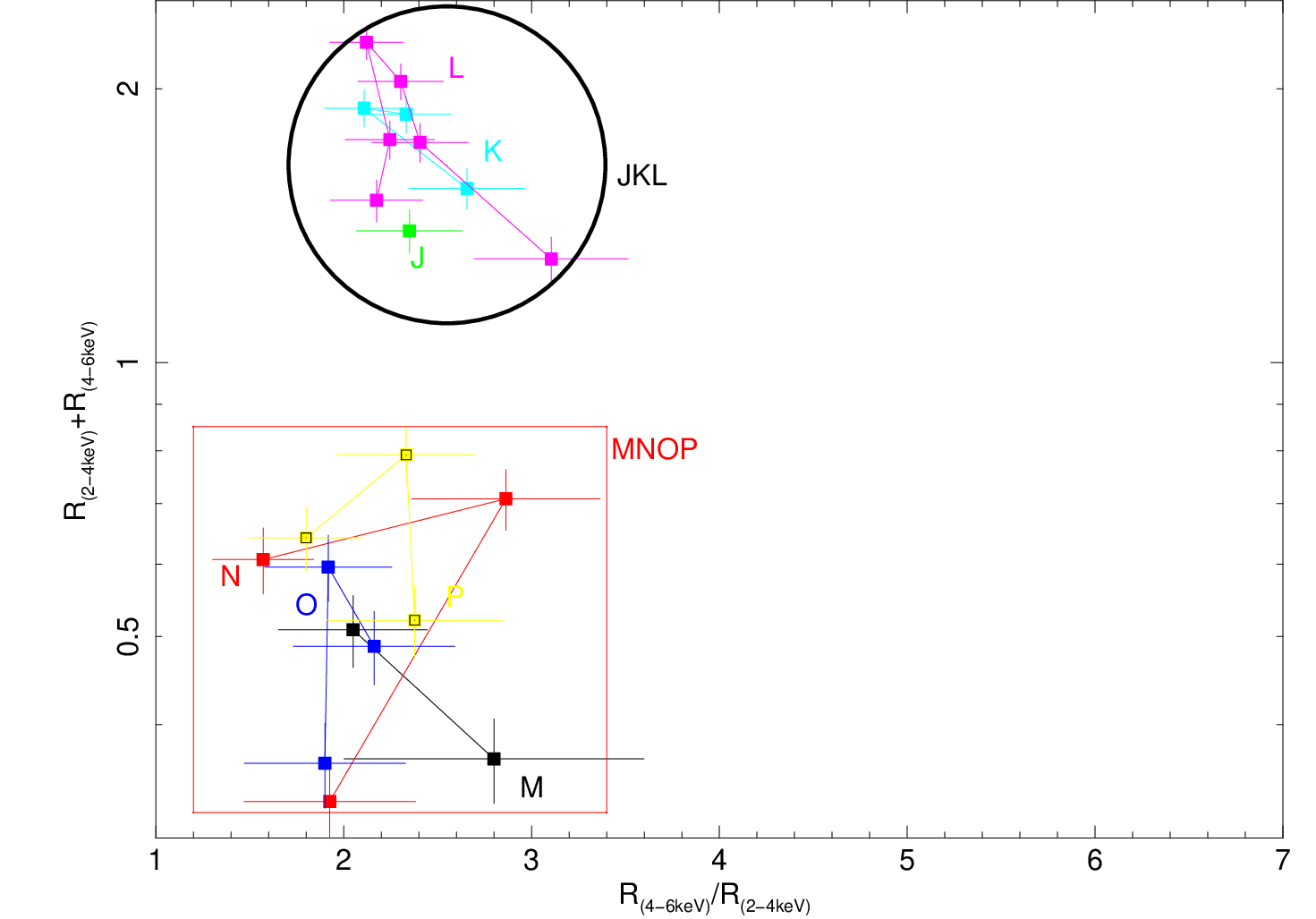} \\
\includegraphics[height=14cm,angle=-90]{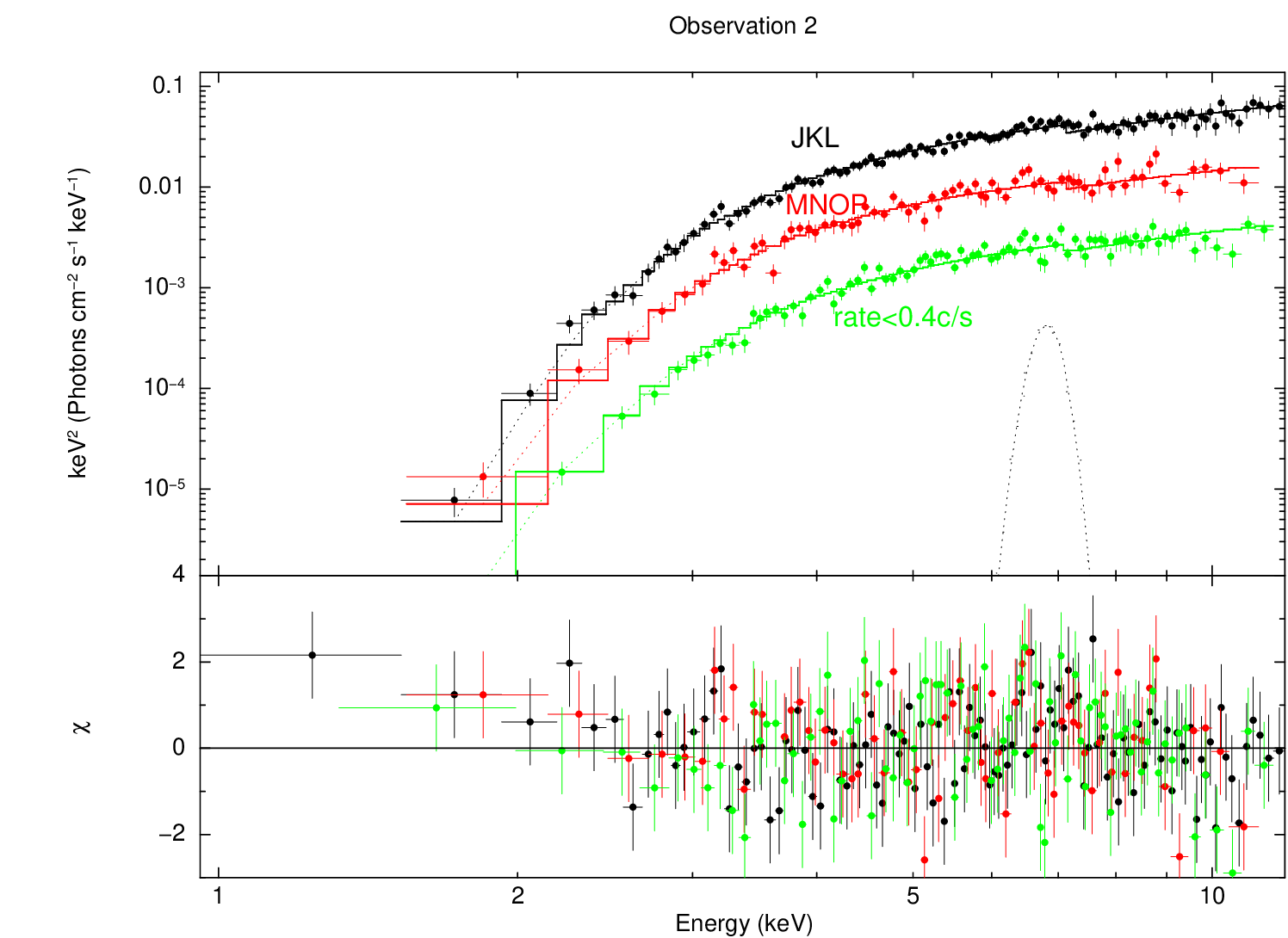} \\
\end{center}
\caption{ Top panel: Hardness ratios $R=Rates_{4-6 keV}/Rates_{2-4 keV}$ for time intervals corresponding to flares displayed in Fig.~\ref{fig:lc3} plotted against the sum $S=Rates_{4-6 keV}/Rates_{2-4 keV}$ for the second observation.
Large circle and square indicate the combined data from flares with the similar flux level and hardness. Bottom panel: unfolded spectra and the model in $E^2f(E)$  and residuals of the following spectral states: MNOP in red, JKL in black, and the state with rate lower than 0.4 c/s in green, fitted with the best model described in the text. 
}
\label{fig:fluxratio2}
\end{figure}

%%%%%%%%%%%%%%%%%%%%%%%%%%%%%%%%%%%%%%%%%%%%%%%%%%%%%%%%%%%%%%%%%%%%%%%%
\begin{figure}
\begin{center}
\includegraphics[height=13cm,angle=-90]{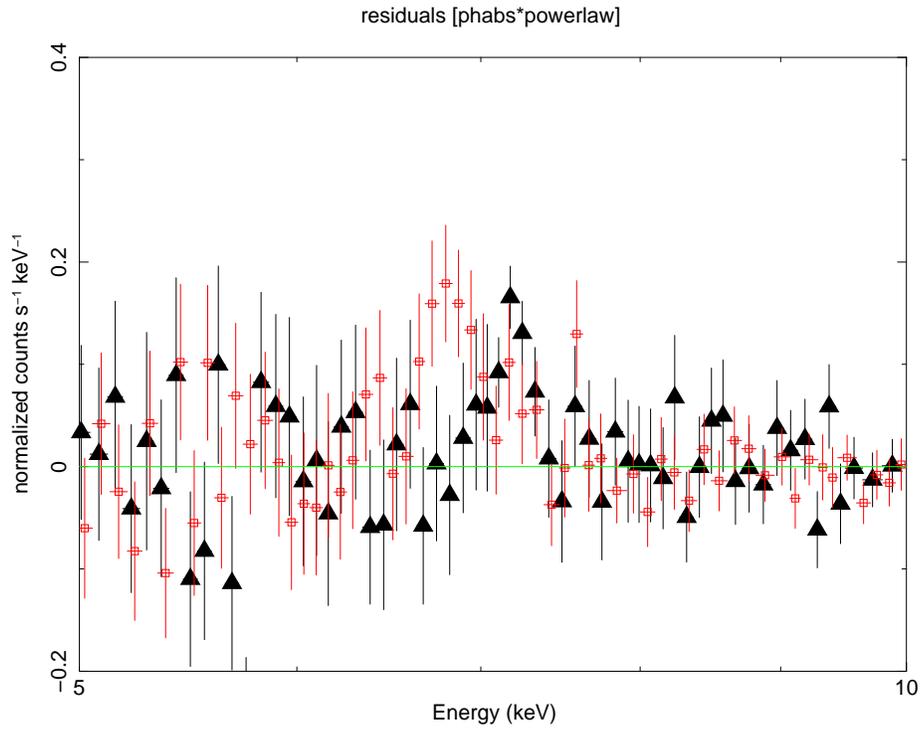} \\
\end{center}
\caption{Residuals in units of count s$^{-1}$ keV$^{-1}$ of the $A_2$ (black triangles) and JKL (red squares) states, in the \xmm\/ first and second observations, respectively.  The used model is an absorbed power law } (see text for details).
\label{fig:res}
\end{figure}

%%%%%%%%%%%%%%%%%%%%%%%%%%%%%%%%%%%%%%%%%%%%%%%%%%%%%%%%%%%%%%%%%%%%%%%%
\begin{figure}
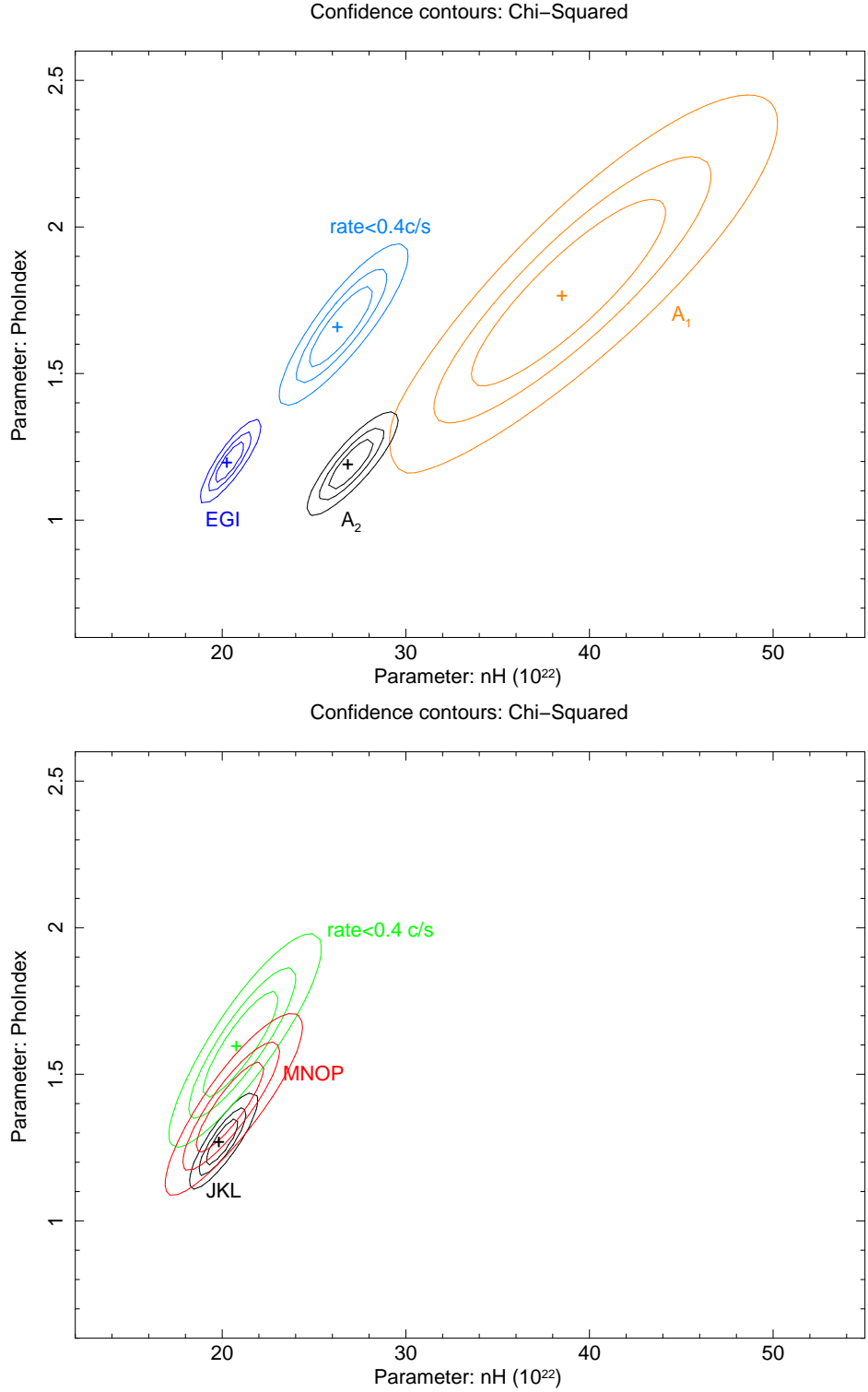

\begin{center}
\includegraphics[height=13cm,angle=-90]{fig6a.eps} \\
\includegraphics[height=13cm,angle=-90]{fig6b.eps} \\
\end{center}
\caption{Top panel:  confidence contours for the best fit parameters to spectral states EGI, A$_1$, A$_2$ and state with rate lower than 0.4 c/s, for the first \xmm\/ observation. 
Bottom panel: confidence contours for the best fit parameters to spectral states JKL, MNOP and state with rate lower than 0.4 c/s, for the second \xmm\/ observation . For both observations all contour are displaying the 68\%, 90\% and 99\% statistical confidence regions.
}
\label{fig:contour}
\end{figure}
%%%%%%%%%%%%%%%%%%%%%%%%%%%%%%%%%%%%%%%%%%%%%%%%%%%%%%%%%%%%%%%%%%%%%%%%

%%%%%%%%%%%%%%%%%%%%%%%%%%%%%%%%%%%%%%%%%%%%%%%%%%%%%%%%%%%%%%%%%%%%%%%%

\begin{figure}
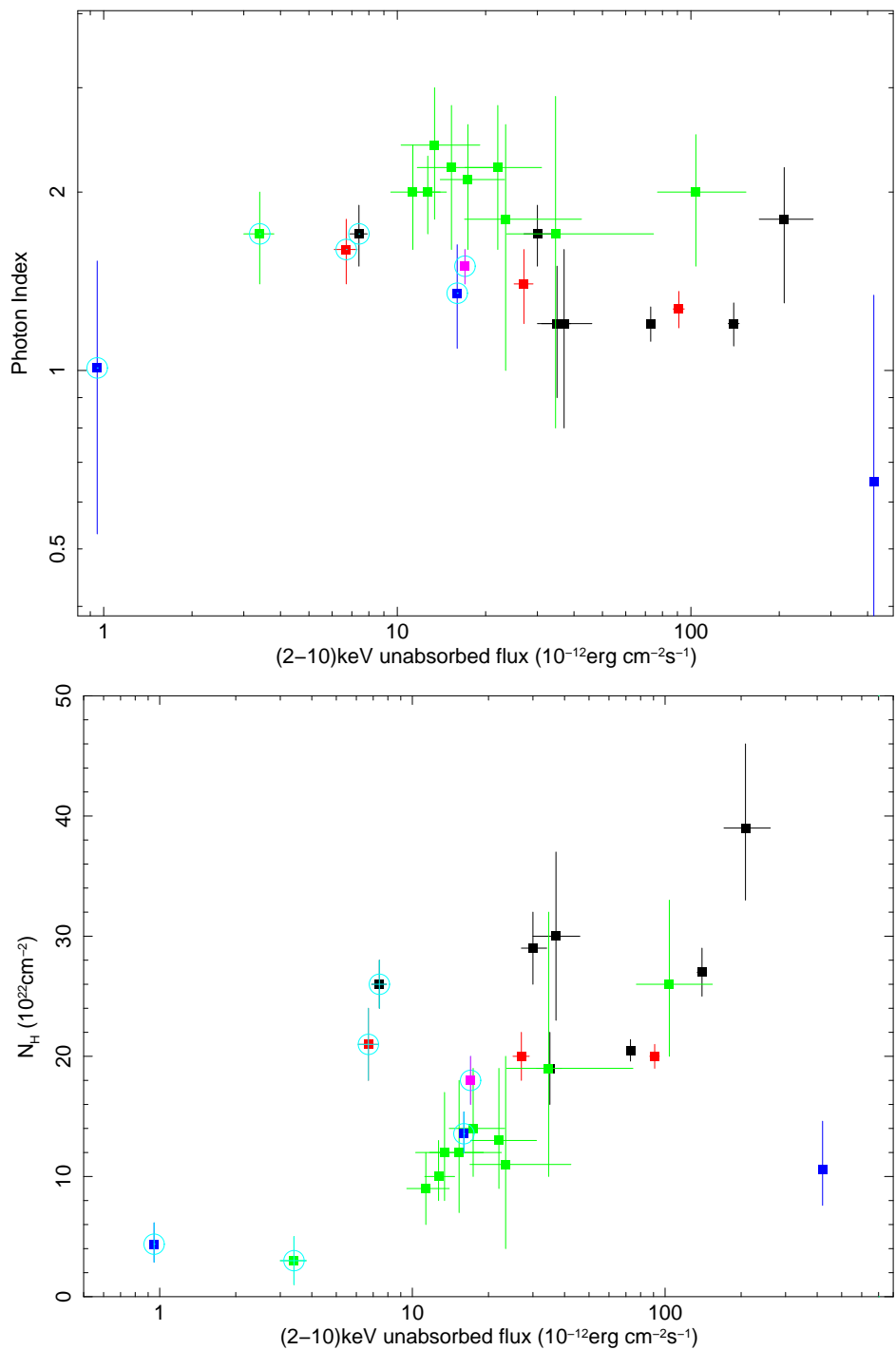

\begin{center}
\includegraphics[height=13cm,angle=-90]{fig7a.eps} \\
\includegraphics[height=13cm,angle=-90]{fig7b.eps} \\
\end{center}
\caption{ Top panel: photon index against unabsorbed fluxes in units of $\rm 10^{-12}erg~cm^{-2} s^{-1}$ in the 2-10 keV energy range. 
Bottom panel: column density in units of $\rm 10^{22} cm^{-2}$  plotted versus unabsorbed flux in units of $\rm 10^{-12}erg~cm^{-2} s^{-1}$ in the 2-10 keV energy range. 
To do this comparison the values from previous observations were computed with the same interstellar abundances used in this paper (Wilms et al. 2000). For blue squares (\sw\/ data from Romano et al. 2013, 2014b)
the flux uncertainties are not available. For details on the colors, see the text.
}
\label{fig:plotPAR}
\end{figure}

%%%%%%%%%%%%%%%%%%%%%%%%%%%%%%%%%%%%%%%%%%%%%%%%%%%%%%%%%%%%%%%%%%%%%%%% Fig 3
\begin{figure}
\begin{center}
\includegraphics[height=13cm,angle=-90]{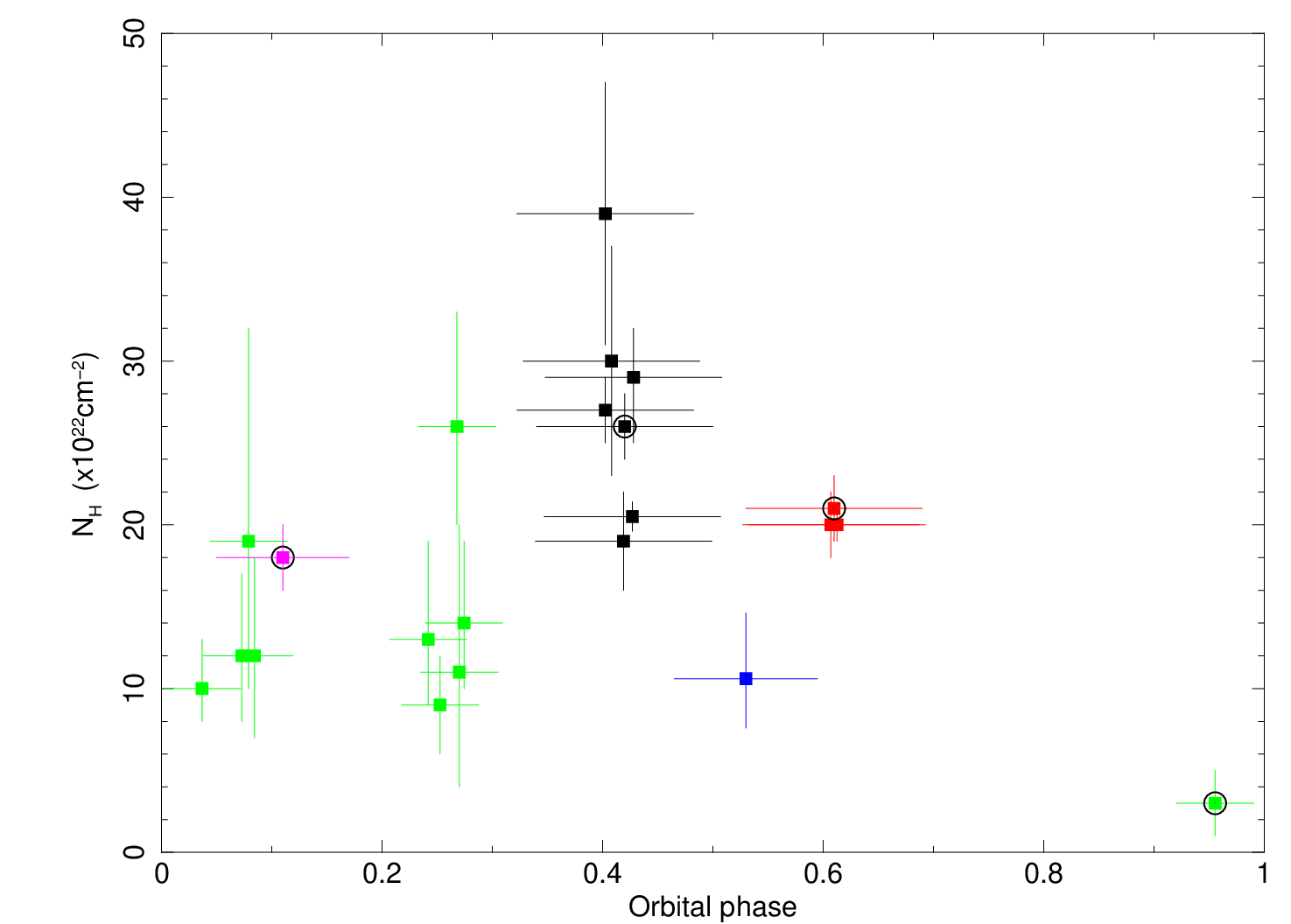} \\
\end{center}
\caption{Column density in units of $10^{22} cm^{-2}$  plotted versus orbital phase.  We display column density values for the first \xmm\/ observation in black, the second \xmm\/ observation in red, the \sax\/ in green (from Fiocchi et al. 2013), \sw\/ XRT in blue (from Romano et al. 2013) and \xmm\/ in magenta (from Bozzo et al. 2012). Square with circles indicate spectral parameters measured during average low emission levels, squares alone indicate parameters measured during an active period.  
}
\label{fig:phase}
\end{figure}
%%%%%%%%%%%%%%%%%%%%%%%%%%%%%%%%%%%%%%%%%%%%%%%%%%%%%%%%%%%%%%%%%%%%%%%%

%%%%%%%%%%%%%%%%%%%%%%%%%%%%%%%%%%%%%%%%%%%%%%%%%%%%%%%%%%%%%%%%%%%%%%%%%%%%%%
%\subsection{Spectral analysis}
%\label{sec:spec}
%%%%%%%%%%%%%%%%%%%%%%%%%%%%%%%%%%%%%%%%%%%%%%%%%%%%%%%%%%%%%%%%%%%%%%%%%%%%%%

Based on the spectral characteristics of this source, including
the column density changes with the flux variations (Fiocchi et al. 2013),
we performed a preliminary analysis for spectral variability by plotting
the hardness versus total rates.
We selected the energy bands 2-4 keV and 4-6 keV and computed the ratio
$R=Rates_{4-6 keV}/Rates_{2-4 keV}$ and the sum $S=Rates_{4-6 keV}+Rates_{2-4 keV}$ of the count rates
using a bin time of 240 s. These energy bands are chosen to investigate the possible column density and flux variations.
We selected eight regions of interest from temporal intervals showing similar flux and similar hardness ratio. 
In  Fig.~\ref{fig:fluxratio1} and ~\ref{fig:fluxratio2} (top panels) the total rate S (intensity) versus the ratio R (colour) for the flares reported
in Fig.~\ref{fig:lc3} are shown, for both observations. Data from flares at similar flux level and similar hardness in the top panels of Fig.~\ref{fig:fluxratio1} and Fig.~\ref{fig:fluxratio2} were 
combined to improve the statistical quality of the 
spectra and better constrain the physical parameters.
In particular, we consider the following regions in the plots of intensity versus colour: a hard region with colour $\geq$ 3.8, a soft region  with colour  $\leq$ 3.8, 
a region with the intensity lower than $\sim$1 c/s and a region with the intensity higher than $\sim$1 c/s.
 Within these colour-intensity regions, for the first \xmm\/ observation we sum data from different flares when the points are superimposed on each others (E+G+I and C+D+H) while we consider single spectral stetes when the flare points
are well separeted (A$_1$, A$_2$, B and F). A similar criterion has been adopted for the second \xmm\/ observation, identifying two spectral states (J+K+L and M+N+O+P).
The different intervals used for the time
selected spectroscopy were indicated with letters in Table 1 and in  Fig.~\ref{fig:fluxratio1} for the first \xmm\/ observation,
and in Table 2 and in  Fig.~\ref{fig:fluxratio2} for the second \xmm\/ observation.
In addition,  spectra of the low  emission level were obtained including events with count rate below 0.4 c/s (the bin time of the light curve is 60 s) 
 corresponding to a net integration time of $\sim$\/15ks  and of $\sim$\/11ks for the first and second  \xmm\/ observations, respectively.
The spectral analysis was performed for both EPIC MOS detectors and EPIC PN detector in the energy range 0.8-12.0 keV.
We show only EPIC-PN spectra in the figures for clarity.\\

The PN and MOS spectra have been fitted simultaneously for each time interval using the model {\sc {phabs*powerlaw}} in {\sc xspec} with the interstellar
abundances of Wilms et al. 2000.
In the case of the spectral states named A$_2$ and JKL 
the residuals show an evidence for a positive excess at iron line energies. For these spectral states a Gaussian component  
was added to the power law model. 
A zoom of residuals in the 5-10 keV band, using a simple absorbed power law model, is shown in Fig.~\ref{fig:res} for the spectral states A$_2$ and  JKL, from the  first and second \xmm\/ observations, respectively. 
For each selected spectral state, the best fit parameters are reported on Table 1 and Table 2 for first and second observations, respectively. \\
Spectra and residuals with respect to the best model are shown in Fig.~\ref{fig:fluxratio1} and Fig.~\ref{fig:fluxratio2} (bottom panels) for the first and second \xmm\/ observation, respectively.\\
For periods in which an emission line is significantly detected, we also report the $\chi^2$ value  obtained using a simple absorbed power law without a Gaussian component (see last row of Table 1 and Table 2).

In Fig.~\ref{fig:contour} (top panel) the confidence contours (68\%, 90\% and 99\% confidence level) for the best fit parameters to spectral states EGI, A$_1$, A$_2$ and ``low state'' %with rate lower than 0.4 c/s 
are shown, for the first \xmm\/ observation. For clarity, we report on the better constrained contour plot only.  In Fig.~\ref{fig:contour} (bottom panel) we show the confidence contours for the best fit 
parameters to spectral states JKL, MNOP and state with rate lower than 0.4 c/s, for the second \xmm\/ observation.

During the first \xmm\/ observation, corresponding to the orbital phase $\sim$ 0.4, the source shows variations in both absorbing column density and photon index. The photon index of the states A$_2$ and EGI is $\sim\/$1.2 at fluxes  
 of $\rm \sim\/10^{-10} erg~s^{-1} cm^{-2}$   and it becomes $\sim\/$1.7 at flux of  $\rm \sim\/7\times10^{-12} erg~s^{-1} cm^{-2}$ (``low state''). 
The spectral parameters of states A$_1$ and B are not well constrained and are compatible with photon indices ranging from 1.1 to 2.5 (at confidence level of 99\%). 
The behavior of the photon index of the spectra A$_2$, EGI and low emission state follows the relation usually observed in accreting X-ray pulsars; X-ray emission is harder when the source is brighter. 
During the first \xmm\/ observation, there is evidence that the column density variation is independent of the unabsorbed flux as shown in Fig.~\ref{fig:contour} (top panel) , 
with different values at high fluxes (states EGI and A$_2$) 
and the same values at different fluxes (states A$_2$ and ``low state'' ). 
During the second \xmm\/ observation, at orbital phase $\sim\/$0.6, both the photon index and the column density remains constant 
(see Fig.~\ref{fig:contour} bottom panel). 

\scriptsize
\begin{table}
\begin{footnotesize}
\centering
\tablewidth{0pt}
\caption[]{Results of the  time selected spectroscopy (letters
mark the same time intervals displayed in Fig.~\ref{fig:lc3}) during the
first \xmm\ observation.
$\Gamma$ is the power law photon index. Unabsorbed flux is in
the 2--10~keV energy range in units of
10$^{-12}$~erg~cm$^{-2}$~s$^{-1}$ and N$_{\rm H}$ is in units of $10^{22}$~cm$^{-2}$. When a Gaussian component was added to the model, E is the centroid in keV, $\sigma$ is the line width in units of keV and EW is the equivalent width in eV.   }
\begin{tabular}{lccccccc}
\hline
\hline
         \noalign {\smallskip}
 \multicolumn{8}{c}{\sc {Observation 1}}\\
\noalign {\smallskip}
\hline
\noalign {\smallskip}
Parameter   	  		&           A1			&A2			&     B		&C+D+H		&F	&E+G+I	& $<$0.4 c/s 	\\
  \noalign {\smallskip}
\hline
\noalign {\smallskip}
Start	(MJD)  		&          56893.832		&56893.835			&    56893.890		&56893.937 (C)	&56894.007	&56893.979 (E)&		\\
 			&           			&			&     		& 56893.968 (D)	&	&56894.049 (G)&		\\
			&           			&			&     		&56894.124 (H)	&	&56894.210 (I)	&		\\
\noalign {\smallskip}
\hline
\noalign {\smallskip}
Exp. PN$^b$ 		&           0.2			&1.7			&     0.6		&1.9		&     0.9	&  3.6 	&     14.6	\\
Exp. MOS1$^b$   		&           0.2			&1.9			&     0.7		&2.1		&     0.9	&  4.0 	&     16.3	\\
Exp. MOS2$^b$   		&           0.2			&1.9			&     0.7		&2.1		&     0.9	&  3.9 	&     16.0	\\
 \noalign {\smallskip}
\hline
\noalign {\smallskip}
N$_{\rm H}$  			&   39$_{-6}^{+7}$     		&$ 27 \pm{2}$		&$ 30 \pm{7}$     	&$ 29\pm{3}$	&$ 19 \pm{3}$    	&$ 20.5 \pm{0.9}$    	&$ 26 \pm{2}$    \\
$\Gamma$                	&  1.8$_{-0.5}^{+0.4}$        	&$1.2\pm{0.1}$		&$1.2 \pm{0.4}$     	&$1.7\pm{0.2}$ 	&$1.2\pm{0.3}$		&$1.20\pm{0.08}$	&$1.7\pm{0.2}$\\
Unabs. Flux             	&   208$_{-37}^{+53}$     	&	$140 \pm{6}$	&37$_{-7}^{+9}$    	&30$_{-3}^{+4}$ &$35\pm{4}$     	&$73\pm{2}$     	&$7.4\pm{0.5}$     \\
E 			& 	...			&	$7.1\pm{0.1}$	&...			&...		&	...		&...	&...\\
$\sigma$ 		& 	...			&		$<$0.2	&...			&...			&	...		&...	&...		\\
EW  			& 	...			&40$_{-20}^{+15}$	&...			&...		&	...			&...	&...	\\
$\chi^{2}$/d.o.f.    	&     ...     		&	261.8/232	&   ...       	&...	&...	      									&...     	&...	\\
$\chi^{2}$/d.o.f. $^a$   	&   79.6/60 		&	298.4/235	&  35.1/39        &83.9/100	&49.0/68	      									&311.4/277		&223.2/186      	\\
\hline
\hline
\label{tab:spec}
\end{tabular}
$^a${$\chi^{2}$/d.o.f.  without Gaussian component.}
$^b${Exposure time in ks.}
\end{footnotesize}
\end{table}
%%%%%%%%%% -------------------------------------------------

\begin{table}
\tablewidth{0pt}
\begin{footnotesize}
\begin{center}
\caption[]{Results of the  time selected spectroscopy (letters
mark the same time intervals displayed in Fig.~\ref{fig:lc3}) during the
second \xmm\ observation.
$\Gamma$ the power law photon index. Unabsorbed flux is in
the 2--10~keV energy range in units of
10$^{-12}$~erg~cm$^{-2}$~s$^{-1}$ and N$_{\rm H}$ is in units of $10^{22}$~cm$^{-2}$. A Gaussian component was added to the model, E is centroids in keV, $\sigma$ is line width in units of keV and EW is equivalent width in eV.   }
\begin{tabular}{lccc}
\hline
\hline
 \noalign {\smallskip}
 \multicolumn{4}{c}{\sc {Observation 2}}\\
\noalign {\smallskip}
\hline
 \noalign {\smallskip}
Parameter   	  		&           J+K+L			&M+N+O+P		& $<$0.4 c/s 	\\
 \noalign {\smallskip}
\hline
\noalign {\smallskip}
Start Time	(IJD)  		&       56895.892   (J) 		&56895.812 (M)			&     	\\
	 		&       56895.923 (K)		&56895.873 (N)			&     	\\
	 		&        56895.967 (L)		&56895.909 (O)			&     	\\
	 		&           			&56895.998 (P)			&     	\\
\noalign {\smallskip}
\hline
\noalign {\smallskip}
Exp. Time PN  $^b$   		&           2.1			&2.3			&     8.5	\\
Exp. Time MOS1    $^b$ 		&           2.3			&2.6			&     9.5	\\
Exp. Time MOS2    $^b$ 		&           2.3			&2.6			&     9.4	\\
 \noalign {\smallskip}
\hline
\noalign {\smallskip}
N$_{\rm H}$  			&  $ 20\pm{1}$     		&$ 20 \pm{2}$&$ 21 \pm{2}$    \\
$\Gamma$                	&  $1.27\pm{0.09}$        	&$1.4\pm{0.2}$		&$1.6\pm{0.2}$\\
Unabs. Flux             	&   $91\pm{4}$     		&	$27 \pm{2}$	&$6.7 \pm{0.6}$     \\
E 		& $6.8\pm0.2$			&...	&...\\
$\sigma$ 		& 	$<$0.6		&	...	&...	\\
EW  		& 	148$_{-90}^{+100}$	&	...	&...	\\
$\chi^{2}$/d.o.f.    	&    216.7/242       		&	...	& ...  	\\
$\chi^{2}$/d.o.f.$^a$    	&    252.9/245       		&	209.1/140	&181.4/142       	\\
\noalign {\smallskip}
\hline
\hline
\label{tab:spec}
\end{tabular}
\end{center}
$^a${$\chi^{2}$/d.o.f.  without Gaussian component.}
$^b${Exposure time in ks.}
\end{footnotesize}
\end{table}
%%%%%%%%%% -------------------------------------------------

\normalsize

 	%%%%%%%%%%%%%%%%%%%%%%%%%%%%%%%%%%%%%%%%%%%%%%%%%%%%%
  	\section{Discussion}
  	%%%%%%%%%%%%%%%%%%%%%%%%%%%%%%%%%%%%%%%%%%%%%%%%%%%%%

The \xmm\/ observations have allowed us to perform an in-depth investigation of the transient source \src\/ at two different orbital phases.
We also followed the source variability in detail, revealing changes in its spectral shape.
The photon index shows significant variations, with values ranging from $\sim$1.2 during high flux intervals (states EGI and A$_2$) to $\sim$1.7 during a low state %the state with count rate lower than 0.4 c/s 
(see Table 1 and Fig.~\ref{fig:contour}, top panel) in the first \xmm\/ observation (at an orbital phase of 0.4). This spectral softening at low luminosity is in agreement with the standard behavior observed in SFXTs. 
 Indeed 
X-ray spectra during very strong flares are usually well described by a flat power law ($\Gamma\sim0-1$) while the photon index increases to values of  $\Gamma\sim1-2$ at lower 
 luminosities of  $\rm \sim10^{33-34} erg~s^{-1}$   (see Romano et al. 2011, 2014b, Sidoli et al 2011). 
Changes to photon index corresponding to changing luminosity are not observed during the second \xmm\/ observation at  orbital phase $\sim\/$0.6.

In Fig.~\ref{fig:plotPAR} we show the spectral index (top panel) and the column densities (bottom panel) against unabsorbed fluxes in the 2-10 keV energy range, using 
spectral analysis of time-selected states reported in Table 1 and Table 2 and archival results. %described in the introduction.
In this way we can track the unabsorbed flux variations by two order of magnitude.
 We show parameter values for the first \xmm\/ observation in black points, for the second \xmm\/ observation in red, for the \sax\/ data in green (from Fiocchi et al. 2013), for the  \sw\/ XRT data in blue (from Romano et al. 2013) 
and \xmm\/ data from Bozzo et al. (2012) in magenta. Since these data all cover similar energy ranges, the derived N$\rm _H$ values should be comparable. Squares with circles indicate spectral parameters measured during average low emission levels, squares alone indicate parameters measured during an active period. 

 The analysis of different flux states confirms changes in the column density,  previously observed using \sax\/ data (Fiocchi et al. 2013) and  highlights the variation of spectral index:
from the top panel of Fig.~\ref{fig:plotPAR} it is clear that the photon index shows significant variations without any clear correlation with unabsorbed flux.
The bottom panel of Fig.~\ref{fig:plotPAR} shows that the N$_H$ values at flux lower than $\rm \sim 10^{-11} erg~s^{-1} cm ^{-2}$ and greater than $\rm \sim4\times 10^{-10} erg~s^{-1} cm ^{-2}$ do not 
confirm the linear correlation between flux and column density observed in this object in the past (Fiocchi et al. 2013), when we consider 
two orders of magnitude in flux. In a restricted range of fluxes, from $\rm \sim 10^{-11} erg~s^{-1} cm ^{-2}$ to $\rm \sim2\times 10^{-10} erg~s^{-1} cm ^{-2}$  this correlation still persists.

To investigate the possible column absorption and the orbital phase correlation, we report in Fig.~\ref{fig:phase} the column density against the orbital phase. 
We display column density values according with colours in Fig.~\ref{fig:plotPAR}.
Squares with circles indicate spectral parameters measured during average low emission levels, squares alone indicate parameters measured during an active period.
Spectral parameters values obtained from spectra with long exposure time (covering $\sim$ one phase) are not included in this plot.

Fig.~\ref{fig:phase} shows that there are significantly higher values of the column density during the active time interval corresponding to the orbital phase of 0.4. 
We note that the average low emission level (squares with circles in  Fig.~\ref{fig:phase}) show a maximum at phase $\sim\/$0.4 and a minimum at phase $\sim$ 0.95.
These data show there could be two levels of the density variations: the first corrisponding to the average low emission states and the second considering the active periods. During the low emission levels (circles of Fig.~\ref{fig:phase}), 
the $\rm N_H$ follows the same behavior that the IBIS intensity (23-50 keV) has versus the orbital phase, with a maximum value at phase $\sim$0.5 and lower values at phases $\sim$0.1 and $\sim$1.0.  
During the active periods, the $\rm N_H$ variations are not correlated with the orbital phase and could indicate changes 
in the accreting material on the neutron star. Obtained $\rm N_H$ values rule out that the observed low emission level (flux lower than $\sim 10^{-11} erg s^{-1} cm ^{-2}$) can be due to obscuration of the emitting region 
by circumstellar material, as infact there are values of the column density during low emission level consistent with the 
ones during the active time intervals (see column density of state with rate lower than 0.4 c/s and A$_2$ state). 
This behavior suggests that the $\rm N_H$ values in the low emission levels could be an indication of the matter distribution along the orbit, while additional mechanisms come into play during flaring activity. \\

The iron fluorescence lines show an interesting evolution: the centroid is at $\sim$6.8 keV when the source is in the JKL state while  shifts up to $\sim$7.1 keV at higher fluxes.
%The fluorescence iron line emission at $\sim$6.8-7.1 keV is complex to explain. 
As the iron line centroid is correlated with 2-10 keV unabsorbed flux (see Table 1 and Table 2),
line emissions at $\sim$ 6.8 keV and $\sim$ 7.1 keV could come from highly ionised iron ions: the ionization level is higher than Fe$_{XXV}$ and  Fe$_{XX}$ for the state A$_2$ and JKL, respectively
(Kallman et al. 2010). 
Since  the theory predicts that  the iron line intensity ratio $\rm I_{K_{\beta}}$/$I_{K_{\alpha}}$ is $\sim0.13~ph~cm^{-2}s^{-1}$ (Kallman et al. 2010), the lack of a strong iron line at 6.4 keV 
during time intervals A$_2$ and JKL exclude that the observed iron line at $\sim$ 6.8-7.1 keV can be fluorescence iron line K$_{\beta}$, not respecting this iron line intensity ratio.
This behavior suggests that the X-ray flux produced by accretion onto the neutron star partly ionized the clump matter. \\
The limited statistics of our \src\/  \xmm\/ data prevent us from studying the expected linear correlations between the  continuum flux and the iron line flux or between the Fe equivalent width and the continuum parameters (N$_H$ and luminosity), 
as reported by Gimenez-Garcia et al. (2015) and Torrejon et al. (2010).

 This work has shown a complex picture that is compatible with accretion from an inhomogeneus wind (in’t Zand 2005,  Walter \& Zurita Heras 2007). The transient emission produced
by accretion of matter from the companion wind indicates change in the wind density and temperature, not clearly correlated with the orbital phase. 
The inhomogeneities in the accreting material are able to give a physical interpretation of the short flares observed in both the \xmm\/ data and the previous \sax\/ ones (Fiocchi et al. 2013).\\
Conversely, the clumpy wind model alone is not able 
to explain the few days long flare observed with \inte\/ from this source (Fiocchi et al. 2010). This evidence confirms that additional mechanisms are nedeed to explain the extreme variability seen in SFXT  (Bozzo et al. 2014, Lutovinov et al. 2013). \\
The two proposed additional mechanisms to inhomogeneus wind are the quasi spherical accretion model (Shakura et al. 2012, 2014) or the centrifugal and/or magnetic gating accretion (Bozzo et al. 2008, 2016).
At this stage, for \src\/ both mechanisms cannot be ruled out, indeed:  \\
1) the theory of wind accretion in HMXB hosting a magnetic neutron star with transitions driven by centrifugal and magnetic barrier (Bozzo et al. 2008, 2016) 
requires an high magnetic field to explain the observed dynamic range (greater than $\sim10^{14} G$):
unfortunately, the magnetic field in \src\/ is unknown. \\
2)the quasi spherical accretion model (Shakura et al. 2012, 2014)  concerns the accretion onto slowly rotating X-ray pulsars: the spin period of \src\/ is unknown. 
Furthermore this theory predicts  two regimes of accretion at the critical  X-ray luminosity value of $\rm \sim4\times10^{36} erg~s^{-1}$. 
The present \xmm\/ data and the previous \sw/XRT results  (Romano et al. 2013) allowed to extend the studied luminosity range, spanning from $\rm \sim6\times10^{33} erg~s^{-1}$ to $\rm \sim3\times10^{36} erg~s^{-1}$. 
Unfortunately, the investigated luminosity values are always lower than critical value  %($\rm \sim4\times10^{36} erg~s^{-1}$) 
preventing to study X-ray behaviour at very high luminosity. 

Finally, we note that the accretion radius and the magnetospheric radius are highly sensitive to variations in the wind velocity and this wind velocity can significantly drop or be completely halted close to
the neutron star when the matter is ionizated (Krticka et al. 2015, Ducci et al. 2010), making the comparison data-model complicated.

%%%%%%%%%%%%%%%%%%%%%%%%%%%%%%%%%%%%%%%%%%%%%%%%%%%%%%%%%
\section*{Acknowledgments}
%%%%%%%%%%%%%%%%%%%%%%%%%%%%%%%%%%%%%%%%%%%%%%%%%%%%%%%%%
The authors acknowledge the ASI financial support via ASI/INAF grants  2013-025-R.0
 and the grant from PRIN/INAF 2014, {\em Towards a unified picture of accretion in High Mass X-Ray Binaries} (PI: Sidoli). \\
We thank Dr. L. Sidoli for suggestions and discussion, improving our work.	\\
We thank the \xmm\/ duty scientists and  science  planners  for  making  these  observations  possible.

%%%%%%%%%%%%%%%%%%%%%%%%%%%%%%%%%%%%%%%%%%%%
\end{document}